\documentclass[apj]{emulateapj}
\shorttitle{Polarization Power Spectra in the CGPS}
\shortauthors{Stutz et al.}

\begin{document}

\title{Power Spectrum Analysis of Polarized Emission from the Canadian Galactic Plane Survey}

\author{R. A.  Stutz\altaffilmark{1}, E. W. Rosolowsky\altaffilmark{1,2}, R. Kothes\altaffilmark{3}, and T.L. Landecker\altaffilmark{3}} 

\altaffiltext{1}{University of British Columbia Okanagan, 3333 University Way,
  Kelowna BC, V1V 1V7, Canada}
\altaffiltext{2}{University of Alberta, Dept. of Physics, 4-183 CCIS, Edmonton, AB, T6G 2E1, Canada}
\altaffiltext{3}{National Research Council Canada, Dominion Radio Astrophysical Observatory, Box 248, Penticton,
  BC V2A 6J9, Canada}

\begin{abstract}
Angular power spectra are calculated and presented for the entirety of the Canadian Galactic Plane Survey polarization dataset at 1.4 GHz covering an area of 1060 deg$^2$. The data analyzed are a combination of data from the 100-m Effelsberg Telescope, the 26-m Telescope at the Dominion Radio Astrophysical Observatory, and the Synthesis Telescope at the Dominion Radio Astrophysical Observatory, allowing all scales to be sampled down to arcminute resolution. The resulting power spectra cover multipoles from $\ell \approx 60$ to $\ell \approx 10^4$ and display both a power-law component at low multipoles and a flattening at high multipoles from point sources.  We fit the power spectrum with a model that accounts for these components and instrumental effects.  The resulting power-law indices are found to have a mode of 2.3, similar to previous results. However, there are significant regional variations in the index, defying attempts to characterize the emission with a single value. The power-law index is found to increase away from the Galactic plane. A transition from small-scale to large-scale structure is evident at $b= 9^{\circ}$, associated with the disk-halo transition in a 15$^{\circ}$ region around $l=108^{\circ}$.  Localized variations in the index are found toward \ion{H}{2} regions and supernova remnants, but the interpretation of these variations is inconclusive.  The power in the polarized emission is anticorrelated with bright thermal emission (traced by H$\alpha$ emission) indicating that the thermal emission depolarizes background synchrotron emission.
\end{abstract}

\section{Introduction}

Magnetic fields permeate the Interstellar Medium (ISM) of the Milky Way.  They carry significant energy \citep{beck11} and play important roles in interstellar processes. On the large scale, magnetic fields somehow mirror the structure of the Galaxy, approximately following the spiral arms \citep{brown10} and they determine the thickness of the disk by supporting it against gravity. Within the arms, on the small scale, they also counter gravity and thereby regulate star formation. They participate in the return of radiation and matter from stars to the ISM through stellar winds and supernova explosions. In supernova blast waves, magnetic fields are responsible for the acceleration of relativistic particles, and the large-scale fields ultimately control the distribution of these particles in the disk and halo.

One of the best tracers of the magnetic field is synchrotron radio emission because it is generated throughout the Galaxy. The polarized fraction of the emission is particularly rich in information, since the polarization angle carries the imprint of the field at the point of origin, and is further affected by Faraday rotation as it propagates through the magnetized ISM. Indeed, Faraday rotation effects dominate the appearance of the polarized sky at decimeter wavelengths.  It should, then, be possible to use images of polarized radio emission to understand the distribution, strength and structure of the Galactic magnetic field. \citet{landecker12} reviews the available data sets, discusses their limitations, and what has been learned to date from radio polarization data about the role of magnetic fields in ISM processes.

There are several approaches to describing the polarized sky.  First, there is the classical approach of searching for objects, or searching for polarization manifestations of objects seen in other tracers (for example \ion{H}{2} regions or stellar-wind bubbles). A second possibility is to search for the products of various depolarization phenomena, differential Faraday rotation (or depth depolarization), or beam depolarization arising from internal or external Faraday dispersion within the ISM \citep{burn66,soko98} and thereby to derive physical conditions from these manifestations. In this paper we take a third approach, using statistical methods to examine the spatial scales of polarization structures.  Such analysis has been done before with focus on calculating the angular power spectra or the structure functions of the polarized emission \citep{tucci00,giar01,tucci02, bruscoli02,giar02,have03,have04,have06, have08,lapo06,carr05,carr10} .  More recently, the work of \citet{gaensler11} demonstrated
that the statistics of the gradient of polarized emission provided clear measurements of the magnetoionic medium.  Structure functions and power spectrum analyses nominally yield equivalent information from the images.  Structure functions are computationally intensive for large maps but can be employed in cases where large fractions of the domain lack data as is often the case for rotation measure data toward point sources.  In contrast, angular power spectra can leverage computationally efficient Fourier analysis but require continuously sampled data.  

Below, we present an angular power spectrum analysis of the continuously sampled polarization maps from the Canadian Galactic Plane Survey.  This investigation presents an unprecedented spatial dynamic range by including both aperture-synthesis and single-antenna data to cover all structural scales from the largest to about one arcminute. Furthermore, the area of sky is large, over 1000 square degrees along the northern Galactic plane. Finally, we adopt a model and processing which explicitly accounts for instrumental effects and the presence of point sources in the data.  

This work expands on the treatment first presented in \citet{stutz-thesis}.  In Section \ref{data} we give a brief description of the data used in this work and outline the technique used in our analysis.  We present maps of the spectral indices in Section \ref{sn:spectral_index_results}.  In Section \ref{discussion} we compare our results with other work of this kind.

\section{Data and Analysis}
\label{data}

\subsection{Canadian Galactic Plane Survey Polarization Maps} 
\label{sn:cgps_introduction} The CGPS polarization data present Stokes $Q$ and
$U$ maps of emission at 1420 MHz from the northern Galactic plane with
an angular resolution of $\sim 1'$ \citep{landecker10}. The survey extends from a Galactic longitude of $l = 66^{\circ}$ to $l = 175^{\circ}$ over the latitude range from $b = -3^{\circ}$ to $b= 5^{\circ}$, an area of 872
square degrees.
An additional high-latitude extension up to $b= 17.5^{\circ}$ was
observed from $l=101^{\circ}$ to $l=116^{\circ}$, covering another
188 square degrees, to make a total area of 1060 square degrees.

The final data products integrate observations from three separate
telescopes and the integration process affects our results.  The
largest angular scales are observed by the DRAO 26-m Telescope, which
has a beam width of $36'$ at 1420 MHz.  The 26-m data are corrected for
ground radiation and are well calibrated providing accurate total
power measurements.  However, the 26-m maps are only sampled at
roughly half the Nyquist rate.  The intermediate angular scales are
observed by the Effelsberg 100-m Telescope, originally observed as
part of the Effelsberg Medium Latitude Survey \citep{reich04}.  The
100-m data sample scales from the size of the observing grid of
$5^{\circ}$ down to the beam with of $9.35'$, providing a link between
the 26-m and the Synthesis Telescope (ST) observations.  However, the Effelsberg data have some uncertainty regarding their absolute calibration.  To integrate
the 26-m (well calibrated) and 100-m data (fully sampled),
\citet{landecker10} perform an image-domain combination of the 26-m
and 100-m data sets and a Fourier-domain combination of the combined
single-dish data and the Synthesis Telescope data.  Data from the Synthesis Telescope are apodized using a
Gaussian bell that reaches 20\% for the longest baselines. The
resulting images have an angular resolution of $\theta = 58'' \csc
\delta$.

There is some overlap in spatial frequencies covered by the telescopes, taken
in pairs.  The DRAO 26-m and Effelsberg data have overlap between
$0.6^{\circ}$ and $2^{\circ}$.  The Effelsberg and DRAO Synthesis Telescope
data overlap between about $15'$ and $40'$.  In each case, amplitudes were
compared in spatially filtered data. \citet{landecker10} noted that
calibration uncertainties between the three components are of the order of
10\%.

\subsection{Calculating Angular Power
  Spectra} \label{sn:supermosaic_aps} 

To determine the distribution of power across the angular scales of
the CGPS polarization data, we calculate and analyze angular power
spectra (APS) of the data. This calculation of APS can also be used
for the single-antenna maps. The method for calculating APS closely
follows that of \citet{have03}, though the analysis differs
significantly due to the wider range of angular scale sampled in the CGPS
data.  

In the analysis that follows, we focus on the polarized intensity data, 
$P=\sqrt{Q^2+U^2}$, where $Q$ and $U$ are the Stokes parameters that describe linear polarization.  Parallel results are found for the $Q$ and $U$ analysis (Section \ref{sn:qucomp}).  An image ($X$), where $X$ may be $Q$, $U$, or $P$, is Fourier transformed into the spatial frequency domain ($\mathcal{V}$) and then multiplied by its complex conjugate to calculate power $\mathcal{P}
 = \mathcal{V}(u,v)\mathcal{V}^{\ast}(u,v)$.  The calculated $\mathcal{P}_X$ is a two-dimensional representation of the power data, so it is averaged in
annular bins of spatial frequency to determine the one-dimensional
behaviour. In each annulus, we also calculate the standard error in
the derived mean as an estimate of the uncertainty of each average.
We express the units of the power spectrum as Jy$^2$ given the
original units of the images are K.  This change in units arises because, when transformed (back) into the Fourier domain,
\begin{equation}
\mathcal{V}(u,v) = \int X(l,b) e^{-2\pi i (u\ell + vb)} dl\, db 
\end{equation}
the integral is carried out over the solid angle of the
image, resulting in units of K sr.  Following radio astronomy
convention, we convert these units into Jy.  In the CGPS polarization
images, a surface brightness of $1.0\,\csc \delta$ K is equivalent to
3.9~mJy~beam$^{-1}$, where the declination dependence arises from the variation of the beam size with declination.  For comparison with previous work, we
report the angular scale in terms of the equivalent multipole $\ell$
that would be derived by a spherical harmonic transform of spherically
gridded data.  A given angular scale $\theta$ is related to multipole
number by $\theta = 180^{\circ}/\sqrt{\ell(\ell+1)}\approx
180^{\circ}/\ell$.

\begin{figure} [h]
  \plotone{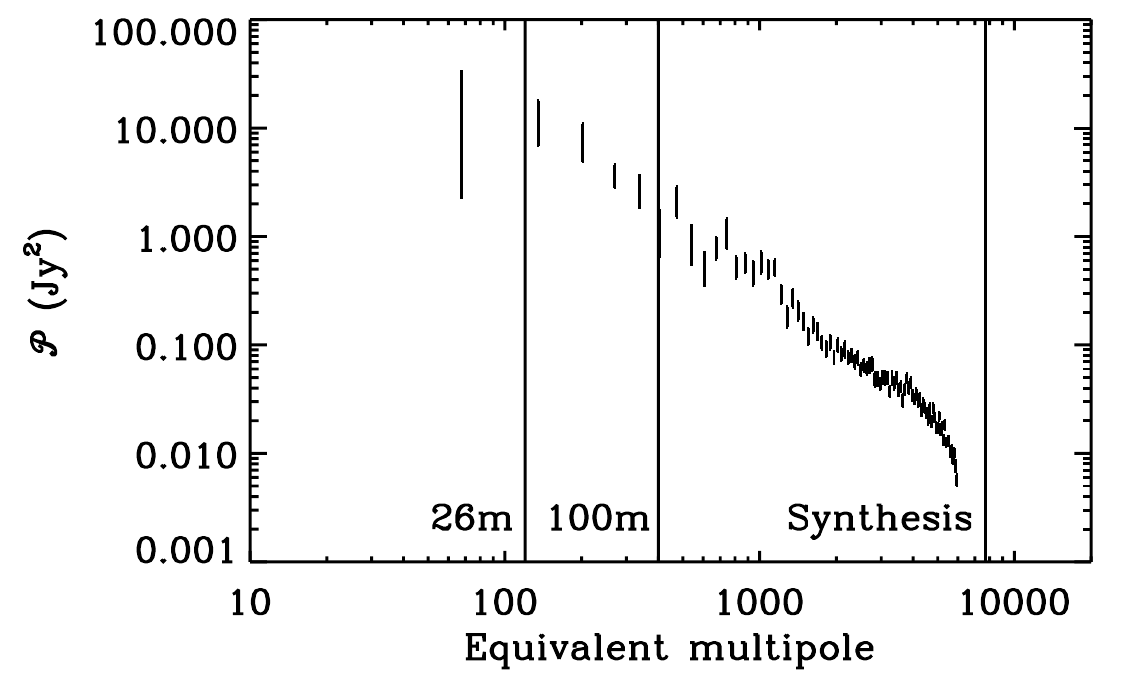}
  \caption{Sample power spectrum of a single $2.67\degr \times
    2.67\degr$ submap centered at l $= 167.9^{\circ}$, $b=
    1.7^{\circ}$. Power-law behaviour is evident at low multipoles,
    as is the effect of the window function at $\ell \gtrsim 3000$. Pixelization    effects are present beyond the support of the window function, but these data are not included in this analysis.}
  \label{fg:sample_ps}
\end{figure}

A representative power spectrum is given in Figure \ref{fg:sample_ps}
highlighting several features discussed later.  The power decreases as
a power law from the smallest sampled multipole to $\ell \sim 10^3$.
If we consider the trend in this portion of the plot as a simple power
law of the form $\mathcal{P}(\ell) \propto \ell^{-\alpha}$, the index
$\alpha$ is usually reported in similar studies e.g.,
\citet{have03}, \citet{giar02}, \citet{carr10}.  Other
structures in the power spectra are not universally seen, though
\citet{have03} and \citet{carr10} do mention a flattening at high multipoles due
to noise in the image.  In particular, \citet{carr10} correct their data for point sources and include a constant offset in their power spectrum regression which accounts for both noise and weak point sources in the data.  Other analyses also account for noise in the data, subtracting the nominal noise power in the data before analyzing the power-law structure \citep{tucci00,bruscoli02}.  

The Gaussian window function (referred to as the ``taper'' in other CGPS papers) applied to the Synthesis Telescope data appears in Figure \ref{fg:sample_ps} from approximately $\ell > 3000$. Power
beyond $\ell>5100$ multipole in Figure \ref{fg:sample_ps} 
probes the pixellization of the CGPS image, so we
restrict our analysis to angular scales of $\ell \leq 5100$.

Figure \ref{fg:sample_ps} shows a feature in the power-law behavior around
$\ell \approx 10^3$.   This may result from the integration of the Effelsberg data with the Synthesis Telescope data.  The typical noise power in the Effelsberg data would be $\sim 0.15\mbox{ Jy}^{2}$ \citep[given a rms surface brightness sensitivity of 8 mK in the Effelsberg $Q,U$ data;][]{landecker10} which indicates significant -- but not strong -- detection of the signal in the regime where the Effelsberg data contribute.  While the Effeslberg data are not expected to contribute significantly at $\ell\approx 10^3$, artifacts of the merger with the Synthesis Telescope may result in the departure from the power-law behavior.  The underlying power-law trend through the data still can be seen across the full range of $\ell$.  

\subsection{Model Fitting to APS} \label{sn:models_alpha} 
To determine accurate spectral index values while accounting for the
flattening and windowing effects at higher multipoles, we fit the power
spectrum with a modified power law function.  We model the flattening
as a constant, non-negative offset.  The Gaussian window function (taper) is applied to the Synthesis Telescope data as a function of baseline length and not $uv$ distance  and is thus represented by an elliptical Gaussian
with major axis $\sigma_{\mathrm{maj}}\equiv \sigma = 5160$ and minor axis $\sigma_{\mathrm{min}}=\sigma \sin\delta$.
Including the effects of annularly averaging an elliptical Gaussian
produces a model for the power spectrum:
\begin{equation}\label{model_eq}
\mathcal{P}(\ell) = \left(A + B\ell^{-\alpha}\right)
\exp\left[-\frac{\ell^2(\sin^2\delta+1)}{2\sigma^2\sin^2\delta}\right]\cdot
I_0\left[-\frac{\ell^2(\sin^2\delta-1)}{2\sigma^2\sin^2\delta}\right].
\end{equation}
The parameters $A$,$B$ and $\alpha$ are free parameters in a least-squares fit and $I_0$ is the modified Bessel function of order 0.  We find much improved stability by also letting $\sigma$ vary in the best fit.  While the value of $\sigma$ is nominally set by the known width for the window function, we attribute the modest variation in the best-fit $\sigma$ to the nonlinear deconvolution and restoration process.  The spectral index $\alpha$ determined by this model should be comparable to those in the literature, which generally fit  multipoles  $\ell \lesssim 1000$.

We calculate the angular power spectrum for subsections of the CGPS to
probe spatial variations in the power spectra.  We extract square $2.67\degr$ submaps from the CGPS data.  For each submap, we
calculate the angular power spectrum and find the best-fitting model.
The analysis is repeated by shifting the extracted region to a
different part of the mosaic.  We oversample our extracted region,
displacing the next extracted region by $2.67^{\circ}/8 = 20'$.  The
submap size limits the range of multipoles probed on the lower end; in
this case submaps probe down to $\ell \approx 67$.  We determine model
parameters for each of the submaps, which we then combine in an image
to determine the parameter trends across the CGPS survey.

\begin{figure}
\plotone{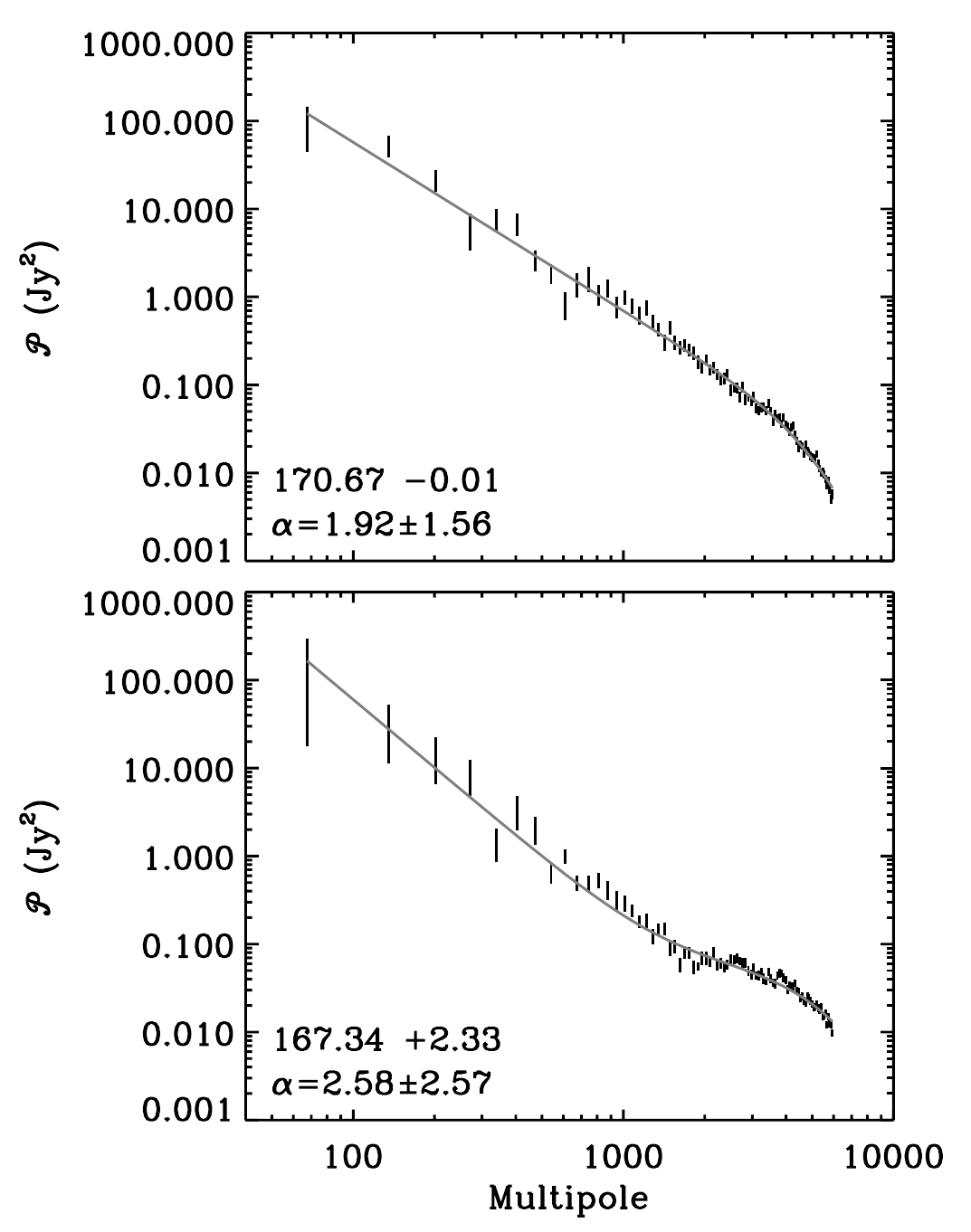}
\caption{\label{sample_ps_model} Sample power spectra and best-fitting
  models for two different parts of the Galactic plane.  The two
  regions, labeled with the respective Galactic coordinates show
  significantly different spectral indices $\alpha$ as indicated in
  the figure. }
\end{figure}

Fits to the power spectra for two single Galactic longitudes are presented in Figure \ref{sample_ps_model}.  The full model, accounting for a variable-width window function appears to reproduce the structure of the power spectrum well.  Covariances between the model parameters are also determined in each fit.  For the fits ultimately used in this study, the average normalized covariance between $\alpha$ and $B$ is 0.995, as is common for slope-intercept pairs of parameters.  The normalized covariance between $\sigma$ and $A$ is $-0.927$, whereas the other covariances are insignificant and vary between 0.5 and 0.7.


\subsection{Point Source Removal}
Point sources on the scale of the Synthesis Telescope beam size are
apparent in the polarized intensity maps, and are found to add
considerable power to the power spectra in which they are included. In
some instances the presence of strong point sources leads to the model
fit failing to converge to a solution, though this appears to be due
to remaining imaging artifacts that were not removed fully in the
cleaning process. Such image artifacts are prevalent especially around
the strongest sources \citep{landecker10}. Point sources are expected
to be background radio galaxies, showing no dependence on Galactic
latitude.  To study the structure of the diffuse polarized emission within our Galactic plane, we remove strong point sources from the map.  

We apply a Roberts edge enhancement filter \citep{handbook-computer-vision} to the polarization maps and any objects in the maps remaining above a threshold of 0.3~K are marked for analysis and possible removal.  We fit
a two-dimensional Gaussian to each object.  If the Gaussian
full-widths at half maximum are less than $2.4'$, the Gaussian fit is
subtracted from the point-like source (the resolution of the maps is
$\sim 1'$). It should be noted that such removal still leaves some
structure in the polarization maps on the size scale of the telescope
beam.  However, more aggressive fitting of the structures did not
yield an improvement in the quality of the results nor did replacing
the region containing the point source with a smooth estimate of the
background.  If the attempt at Gaussian fitting fails, the object
remains in the polarization maps.  Thus bright, moderately resolved
sources will remain in the maps. 

\begin{figure}
\plotone{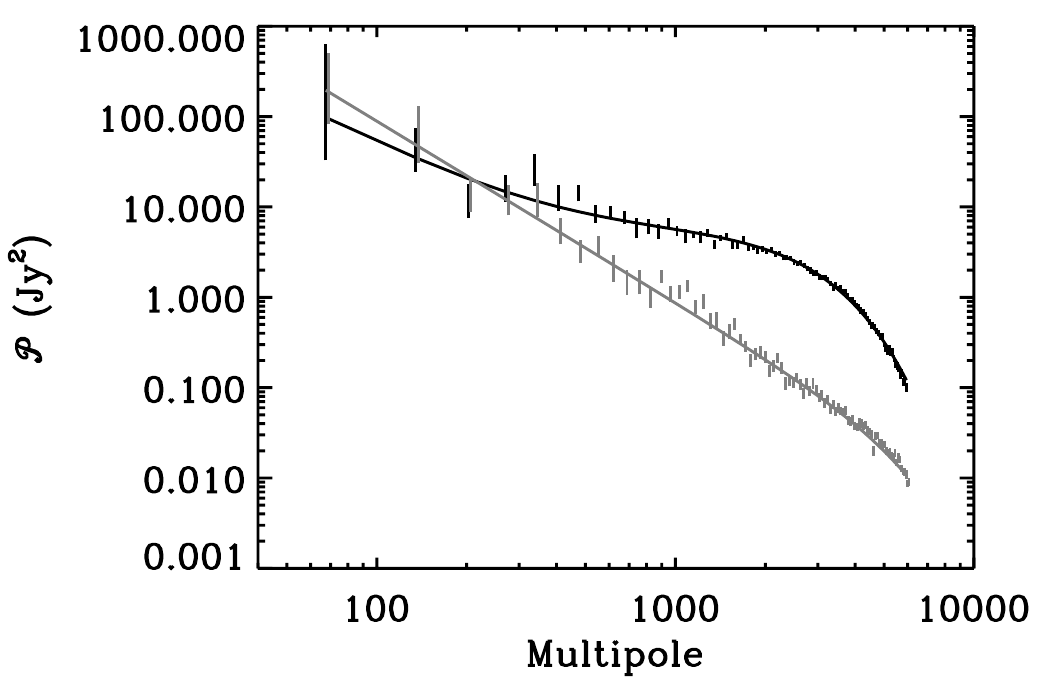}
  \caption{Angular power spectrum for a submap centered at
    $l=167.6^{\circ}$, $b=-0.7^{\circ}$ for maps without (black) and
    with (gray) point-source removal. The submap of this region
    contains significant power in point sources. Point source removal
    yields a steeper slope to the angular power spectrum.}
  \label{fg:ptsub1}
\end{figure}

Figure \ref{fg:ptsub1} shows a comparison of angular power spectra
before and after point source removal. In general, the model
parameters for the regions containing point sources are altered by the
removal of the point sources: the constant power offset parameter $A$
is reduced while the spectral index $\alpha$ is increased.  The reduction of $A$ is expected since the removal of the point sources significantly
lowers the total power in the map.  Since the power spectrum of a
point source is flat, the subtraction has a larger relative impact at
high multipoles where the power is lower. The reduction in small scale
structure by removal of the point sources steepens the spectral index $\alpha$ of the model (Equation \ref{model_eq}).  This change in the spectral index $\alpha$ arises because of the negative covariance between the index and $A$ (Section \ref{sn:models_alpha}).  

\section{Results} \label{sn:spectral_index_results} 

\begin{figure*}
\stepcounter{figure}
\figurenum{\arabic{figure}a}
\plotone{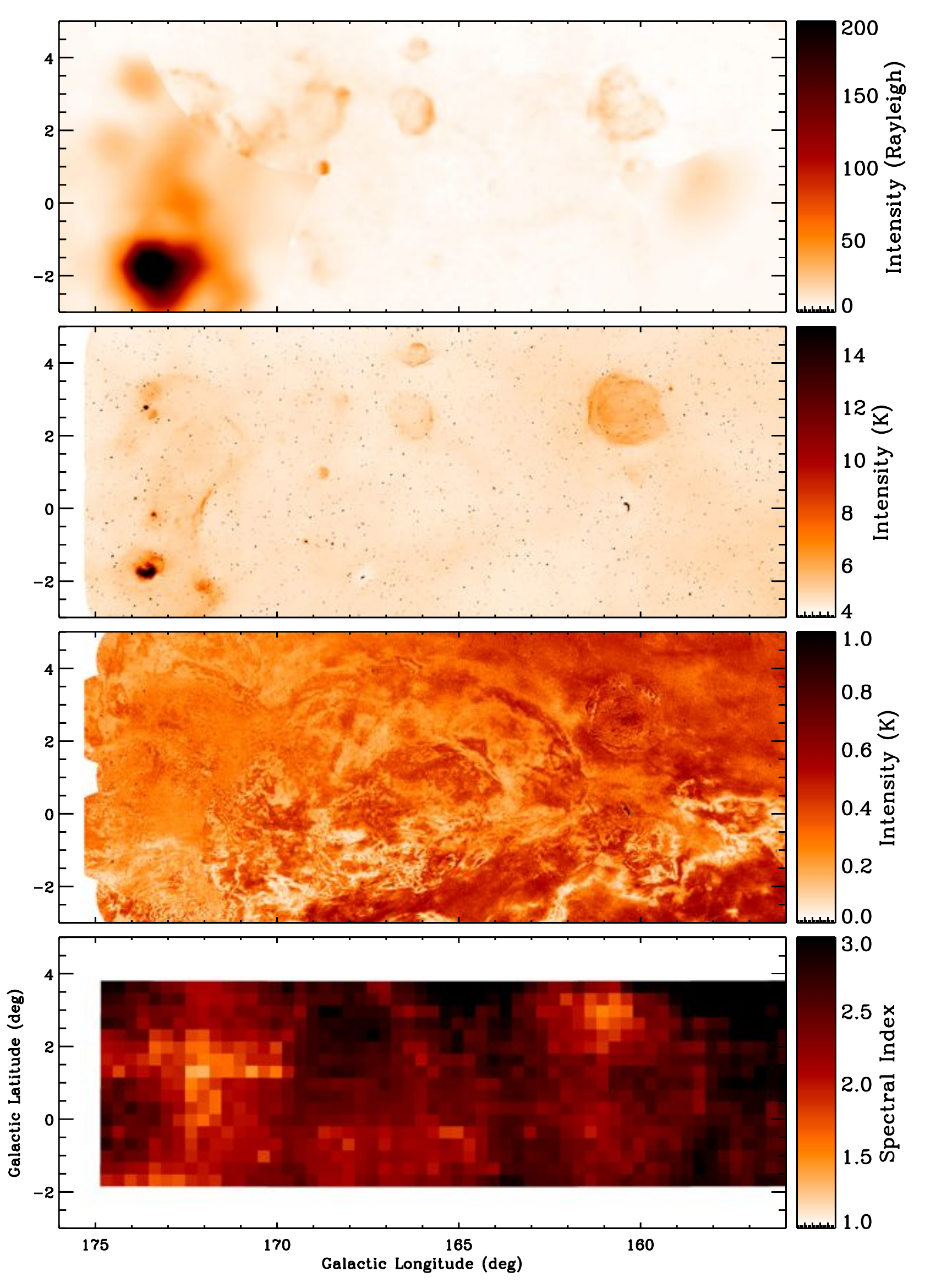}
\caption{Maps of H$\alpha$ emission (top) total intensity $I$ (second), polarized intensity $P$ (third), and spectral index $\alpha$ (bottom) for $l=157^{\circ}$ to
  $l=175^{\circ}$. }
\label{fg:maps}
\end{figure*}

\begin{figure*}
\figurenum{\arabic{figure}b}
\plotone{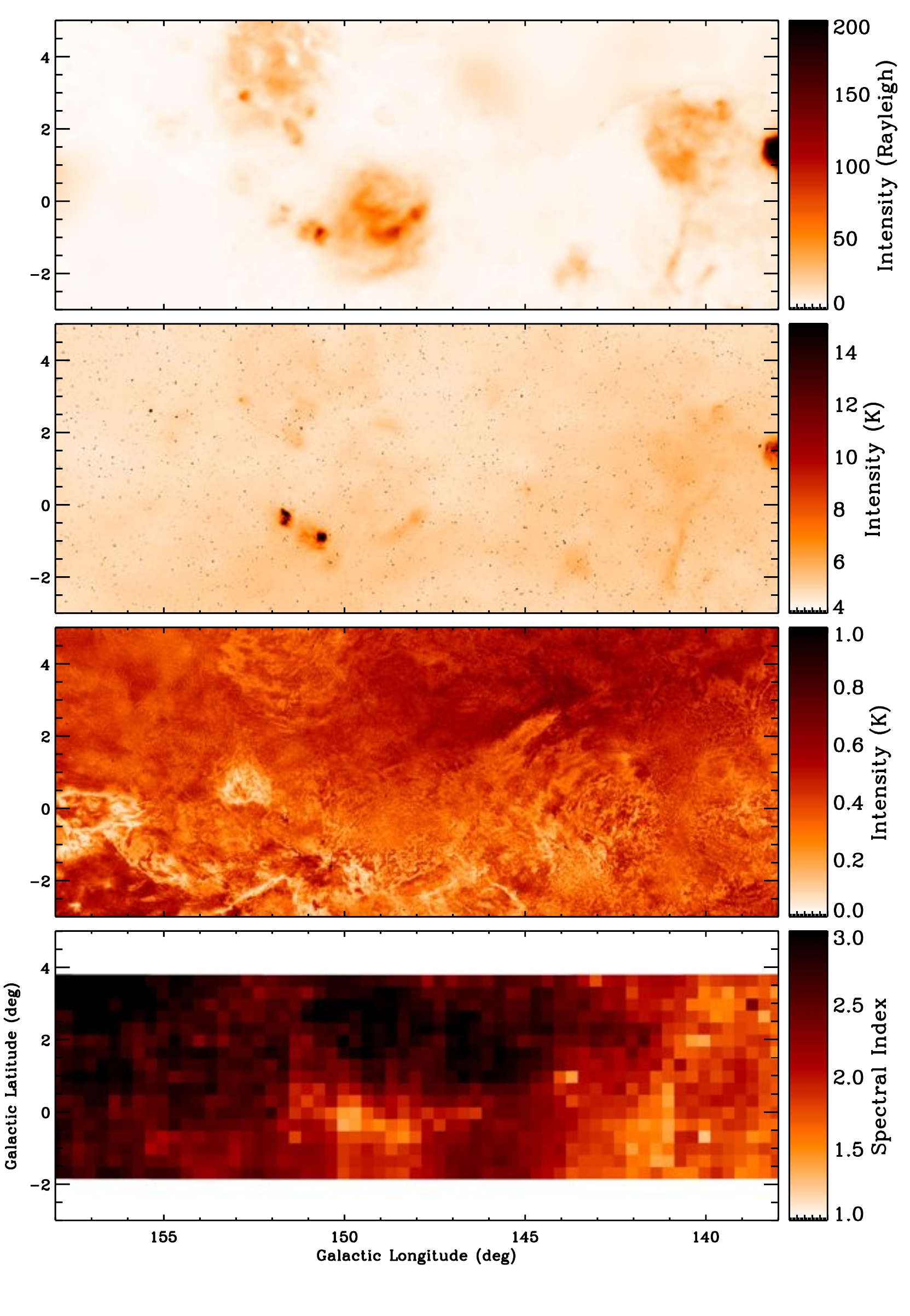}
\caption{As Figure \arabic{figure}a but for $l = 139^{\circ}$ to
  $l = 157^{\circ}$.}
\label{fg:mapsb}
\end{figure*}

\begin{figure*}
\figurenum{\arabic{figure}c}
\plotone{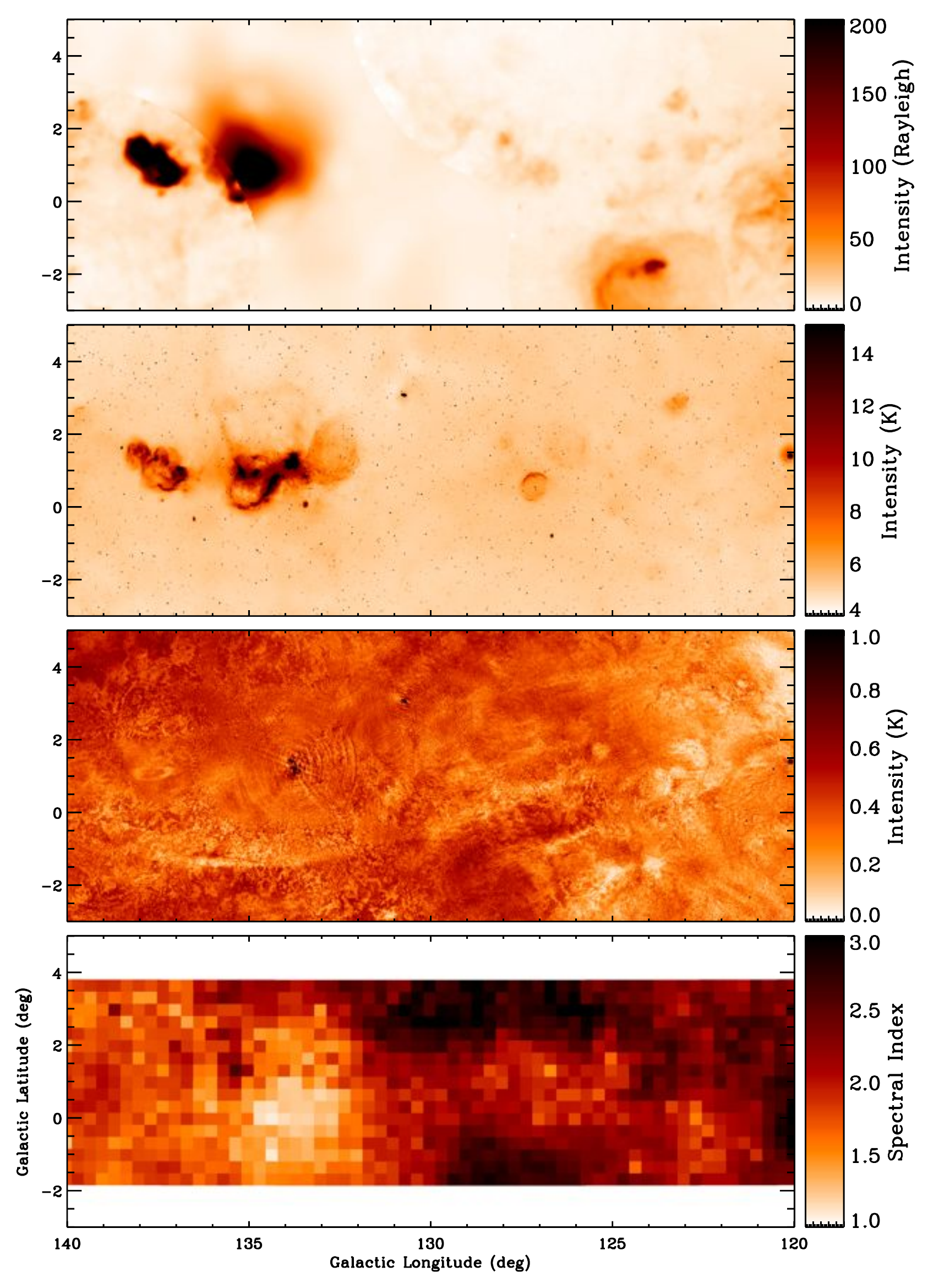}
\caption{As Figure \arabic{figure}a but for $l=121^{\circ}$ to $l=139^{\circ}$.}
\label{fg:mapsc}
\end{figure*}

\begin{figure*}
\figurenum{\arabic{figure}d}
\plotone{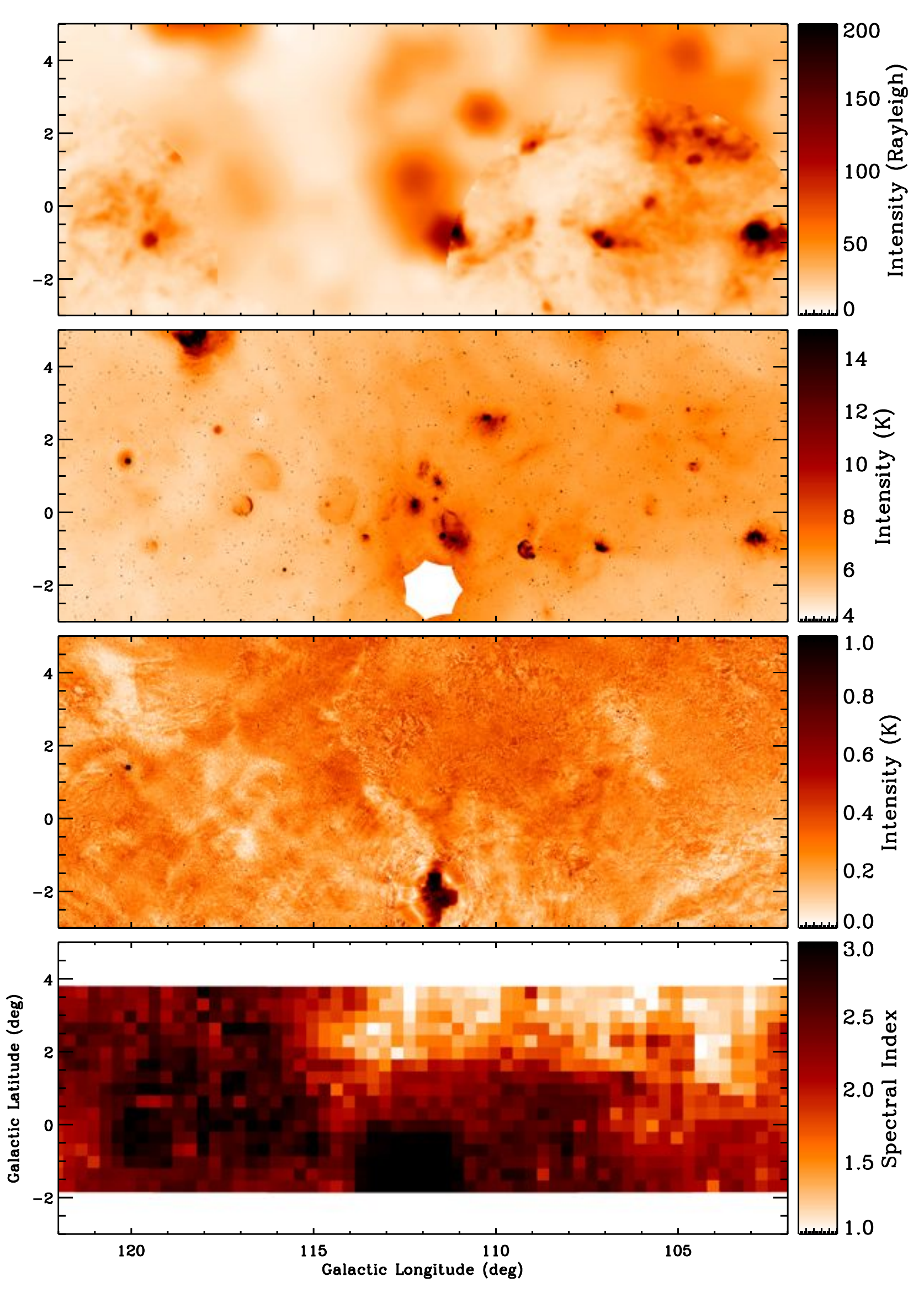}
\caption{As Figure \arabic{figure}a but for $l=103^{\circ}$ to $l=121^{\circ}$}
\label{fg:mapsd}
\end{figure*}

\begin{figure*}
\figurenum{\arabic{figure}e}
\plotone{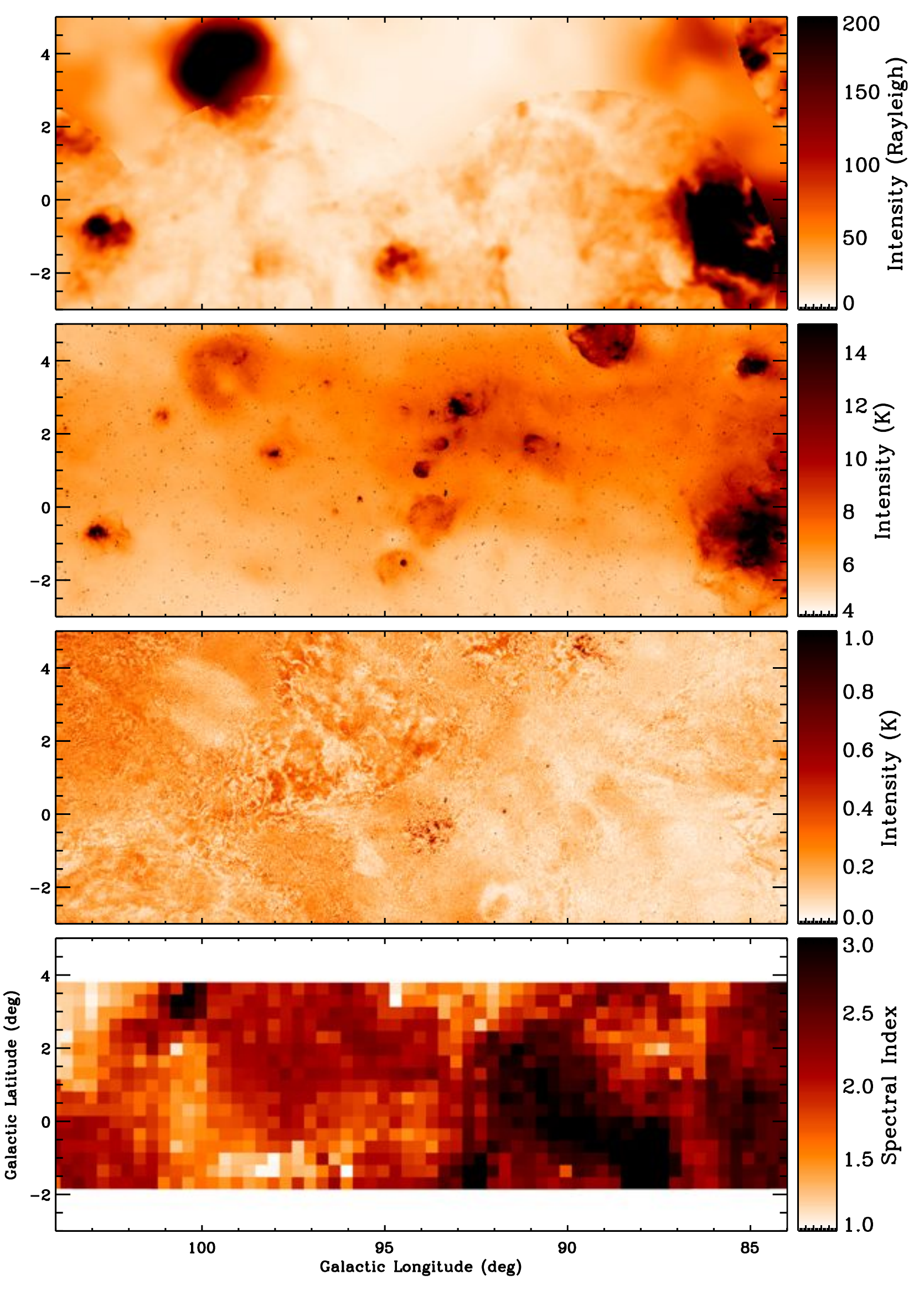}
\caption{As Figure \arabic{figure}a but for $l=85^{\circ}$ to $l=103^{\circ}$}
\label{fg:mapse}
\end{figure*}

\begin{figure*}
\figurenum{\arabic{figure}f}
\plotone{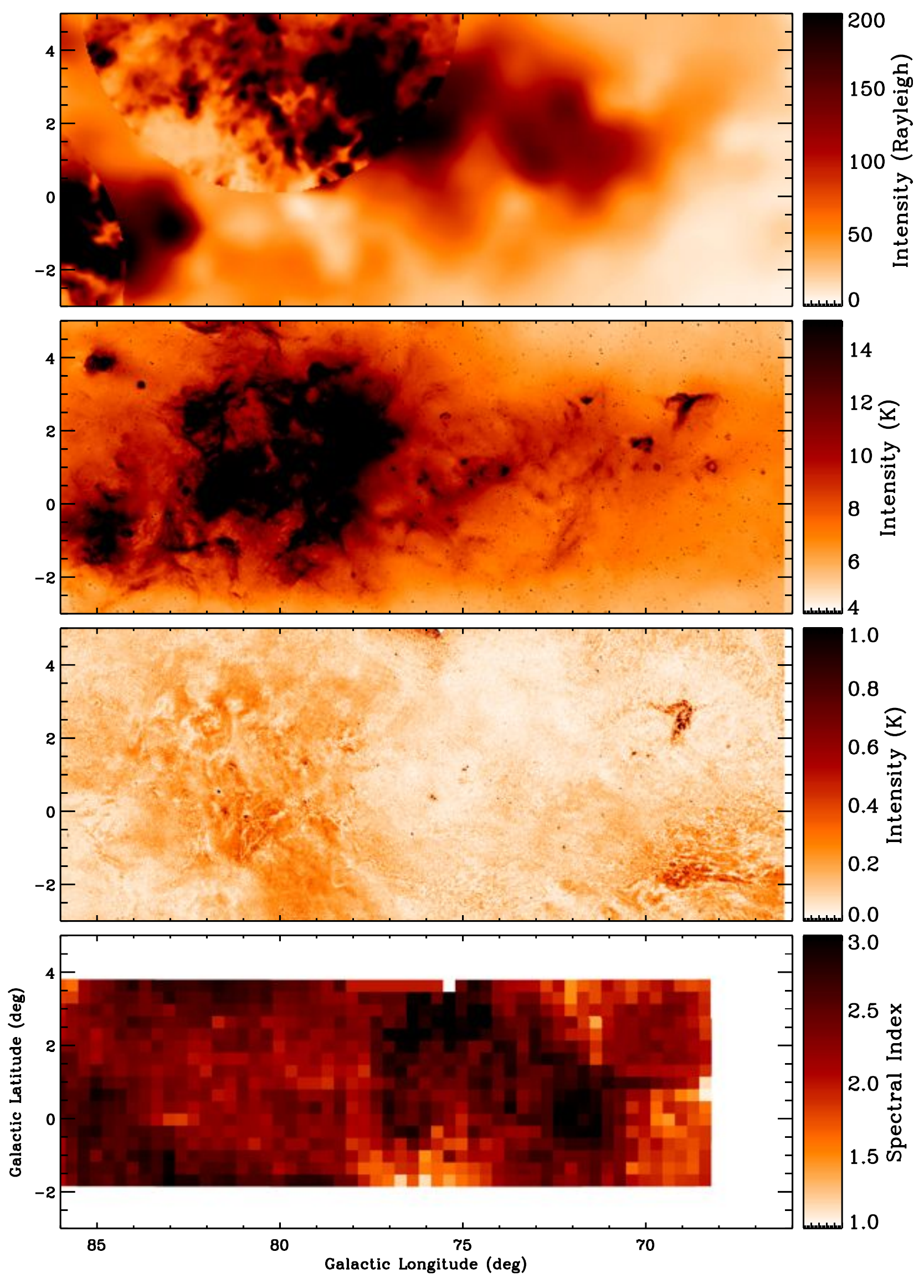}
\caption{As Figure \arabic{figure}a but for $l=67^{\circ}$ to $l=85^{\circ}$}
\label{fg:mapsend}
\end{figure*}

\begin{figure}
\plotone{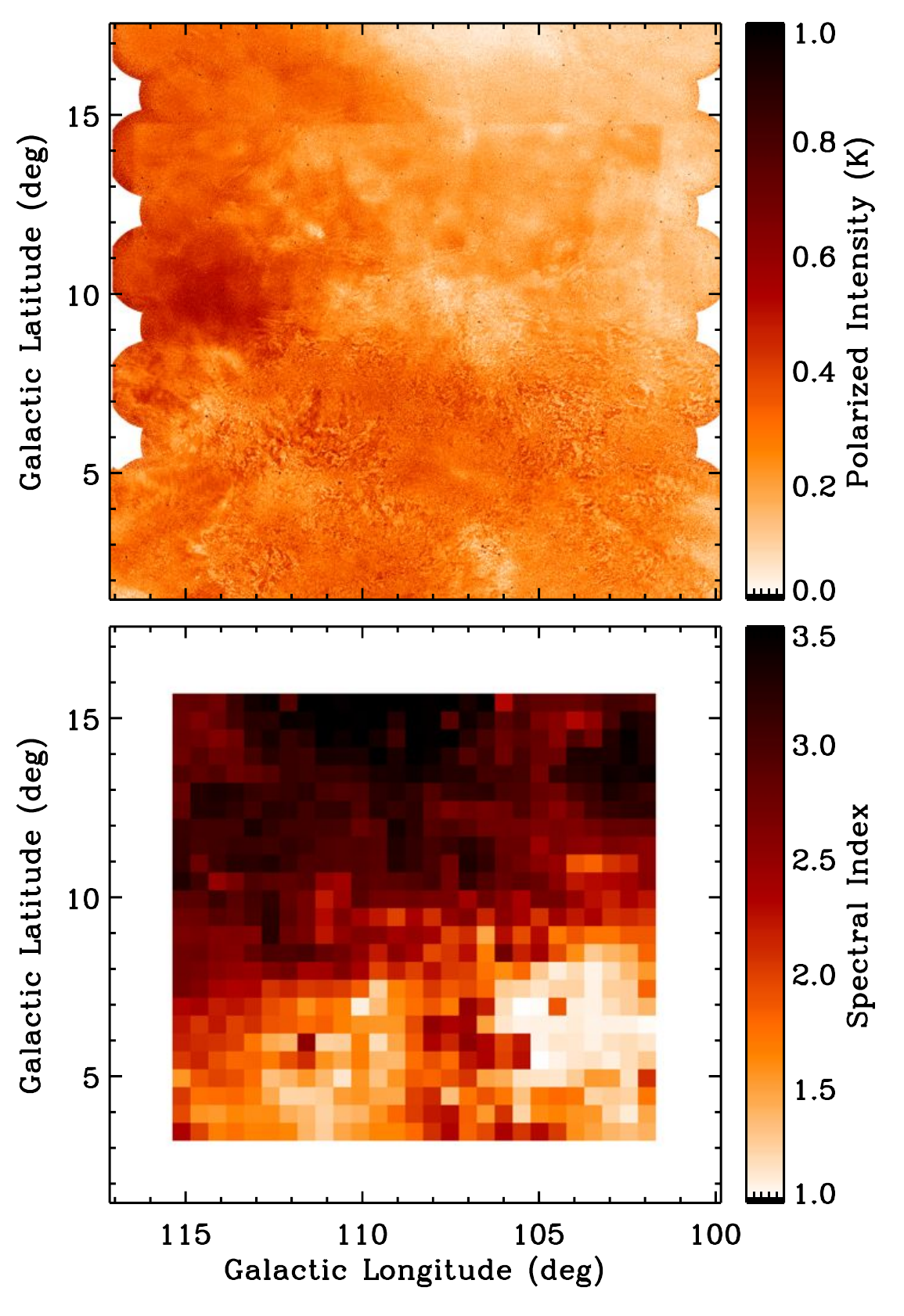}
\caption{Polarized intenity ($P$, top) and spectral index ($\alpha$,
  bottom) maps for the high-latitude extension to the CGPS.}
\label{fg:highlat}
\end{figure}

The results of the spectral index fitting are shown in Figures \ref{fg:maps}
through \ref{fg:mapsend} and Figure \ref{fg:highlat}.  We sample maps of spectral index $\alpha$ by
plotting the value of the spectral index at the center of every submap
extracted.  For comparison we plot H$\alpha$ emission \citep{fink03}, total intensity maps $I$ and the polarized intensity maps $P$ for each region examined.  Some information at the top and bottom edges of the maps is lost due to the finite size of the
submaps. Smaller submaps would allow a more thorough sampling toward the edges
of the maps, but the smaller size would also limit the multipole ranges in
each power spectrum. In the maps of spectral indices, darker pixels indicate
steeper power spectra, and thus excess power on large scales. Lighter pixels
indicate regions with significantly more power on smaller scales. Blanked
regions denote a failure of the model fitting procedure to converge to a
reasonable solution.   The number of fits that fail owing to poor model fits  dramatically decreases after point source subtraction.  The corrupted data are confined to a small region with residual imaging artifacts around Cassiopeia A.  Those data are suppressed in the following analysis.  

\section{Discussion} 
\label{discussion}
\subsection{Understanding Flat Power Contributions} \label{sn:flat_models} 

The flat component observed in the angular power spectra could result from
several features. \citet{have03} observes the flattening in power
spectra of Stokes $Q$ maps, and attributes the flattening to noise.
They truncate the fitting at lower multipoles to avoid including its
effects in the spectral index calculations.  \citet{carr10} also note flattening and attribute it to point sources and noise contributions in the APS and use point-source removal to correct their data.
In the CGPS data, the flattening dominates the synthesis multipoles, which range from $\ell \sim 400$ to $\ell \sim 8000$.  In this section, we establish the possible sources of flattening in the angular power spectrum and the results of our image processing. 

\begin{figure}
  \plotone{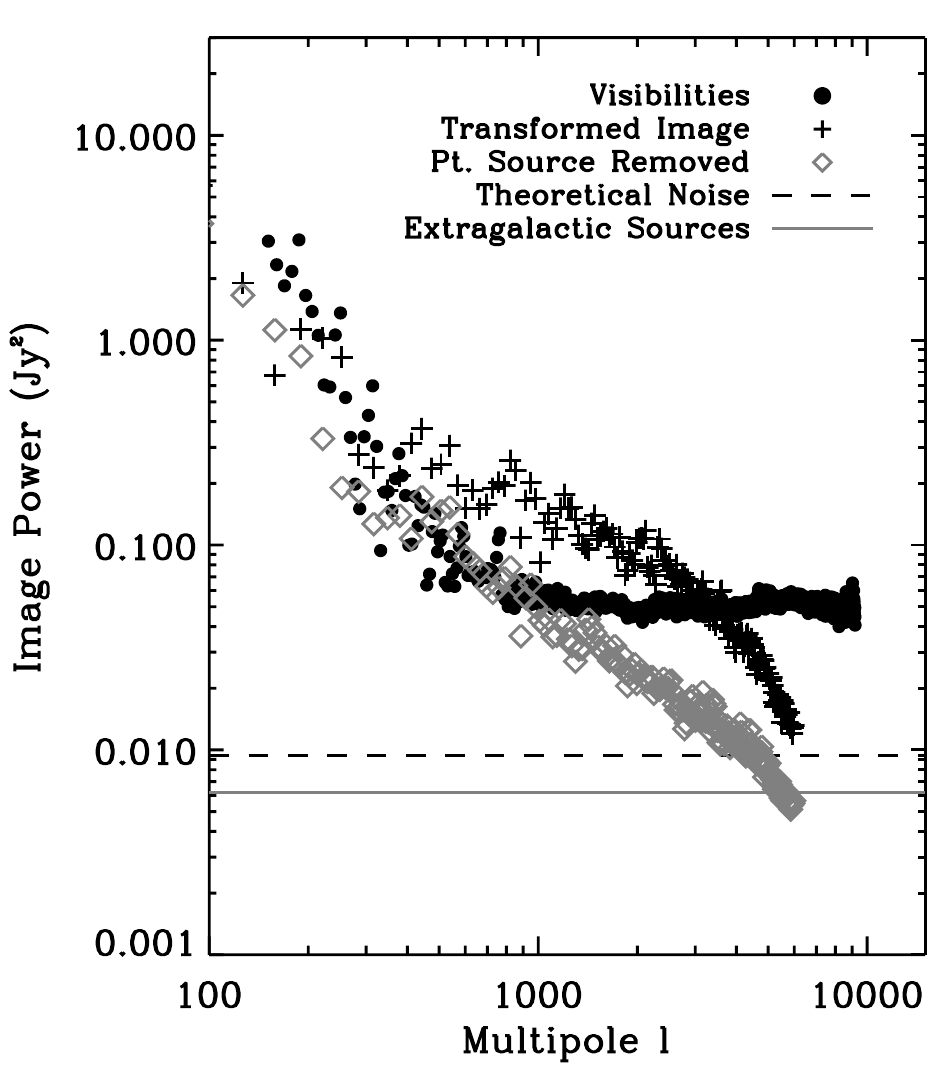}
  \caption{Power spectrum (filled circles) of a single Synthesis Telescope field calculated directly from visibility data at $l=91.6^{\circ},b=1.0^{\circ}$.  The horizontal lines show the contribution from noise power based on direct calculation.  The noise estimates are significantly below the flat portion of the power spectrum, indicating a significant contribution from point sources.  The black crosses and gray diamonds show the transformed image from the CGPS image for the corresponding field before and after point source subtraction respectively.  These results show that the flattening seen in some power spectra is due to point sources in the data and that subtraction mitigates the effects.  The power expected due to faint extragalactic point sources is also indicated and is significantly below the power spectrum. }
  \label{fg:synth_only_ps}
\end{figure}

A power spectrum from a field observed with the Synthesis Telescope
alone is shown in Figure \ref{fg:synth_only_ps}.  This power spectrum
is calculated from the power in the visibility data alone, avoiding
imaging features that arise in transforming to and from the image
domain.  The flattening is apparent out to the higher multipoles than
in the image data because the window function has not been applied.  The extent
of the flat portion of the power spectrum is seen to cover most of the
signal sampled by the Synthesis Telescope.  There is some disagreement between the visibility data for a single field and the transformed CGPS image data.  This discrepancy arises because the CGPS image data contain information from the Effelsberg telescope and a given spatial position appears in multiple CGPS fields.  We focus on using the transformed image data because the derived APS parameters agree well with the visibility data alone and the inclusion of the Effelsberg and 26-m data allow the results to be extended to lower multipoles.  

A flat power spectrum is indicative of either a white noise power
distribution or the contributions from point sources.  The noise level
in an individual CGPS field is  $\sigma_{\mathrm{map}} = 0.2$ mJy beam$^{-1}$, in \citet{landecker10}, which transforms into a noise power per multipole of $\mathcal{P}_{noise}=2N_{beams}N_{pixel}\sigma_{\mathrm{map}}^2$ where
the factor of 2 accounts for the noise in the $Q$ and $U$ images and
the $N_{beams}$ term accounts for the stochastic contributions of
noise power $=(\sqrt{N})^2$.  Similarly, the number of pixels per beam, $N_{pixel}$, is included to account for the oversampling of the beam, contributing to the total noise power.   For single fields of the CGPS, $\mathcal{P}_{noise} \sim 0.0015\mbox{ Jy}^{2}\csc\delta$.  

A final possible contribution to the flattening is the contribution of a population of point sources (e.g., extragalactic background galaxies) that are too faint to be identified as individual objects for point-source subtraction.  As a limit on this component, we use the data from \citet{drao-deep-pol}, who made deep observations of the ELAIS N1 field with the Synthesis Telescope.  They obtained a source count distribution of the extragalactic polarized point source population.  Using their catalog, we calculated the average power that would be expected in a single CGPS pointing.  The level is indicated in Figure \ref{fg:synth_only_ps} and is lower than the theoretical noise floor.  The expected power from noise or faint (extragalactic) point sources is significantly lower than the observed angular power spectrum. However, once the power spectrum is corrected for the presence of bright point sources, the overall amplitude of the power spectrum is significantly decreased and the derived offset ($A$ in Equation \ref{model_eq}) is consistent with expected noise levels. We conclude, based on this analysis, that the flattening in the CGPS power spectra results almost entirely from bright point sources and that the point-source-removal algorithm is effective at extracting most of the power contributions from these sources.  

We note that the complete model appears to be sufficient for describing the angular power spectra of the CGPS data.  For all spectra so fit, we do not see any evidence for breaks in the power spectra that cannot be attributed to point sources.  For example, we do not see a second power-law regime that is revealed when the power spectra are point-source corrected.   In particular, this lack of additional structure in the power spectrum indicates that we have detected no transitions between turbulent regimes as have been argued for based on rotation measure structure functions \citep{have08}.  Those transitions are inferred to occur at small scales (a few pc) and would thus be not well probed by our data.  The CGPS angular resolution of $1'$ projects to this distance in the Perseus arm (2-3 kpc), which is roughly the distance of the polarization horizon toward the outer galaxy \citep[Section \ref{sn:features};][]{uyaniker03}.  Any effect would be present only in the highest multipoles in the data.  While we see no clear indication of such an effect, these data are most affected by the window function and the flat power contributions.

\subsection{Statistics of the Power Spectra}

The slope of the power spectrum describes the relative distribution between
structure (power) on large and small scales.  Large values of the spectral
index ($\alpha > 3$) indicate significant large-scale structure in the
polarized emission.  A flatter power spectrum ($\alpha < 2$) shows significant
small scale structure, whereas $\alpha=2$ would represent equal power on all spatial scales.  A principal result of this analysis is that the shape of the power spectrum changes significantly across the CGPS.  

The least-squares fitting provides an estimate of the uncertainty of the
derived parameters which can be compared to the distribution of the
parameters.  Uncertainties in the individual data composing the
observed power spectrum are deduced from the scatter of the measurements in the
annular averages.  The uncertainties are taken as the standard error in the
mean.  The uncertainties in the parameters are derived from the least-squares
optimization and accounting for covariance between the fit parameters.  The
derived fit parameters are well determined.  For example, the typical
uncertainty on the index $\alpha$ is $\langle\sigma_\alpha\rangle = 0.15$ vs. a
typical magnitude of $\langle \alpha \rangle = 2.3$.  

\begin{figure}
\plotone{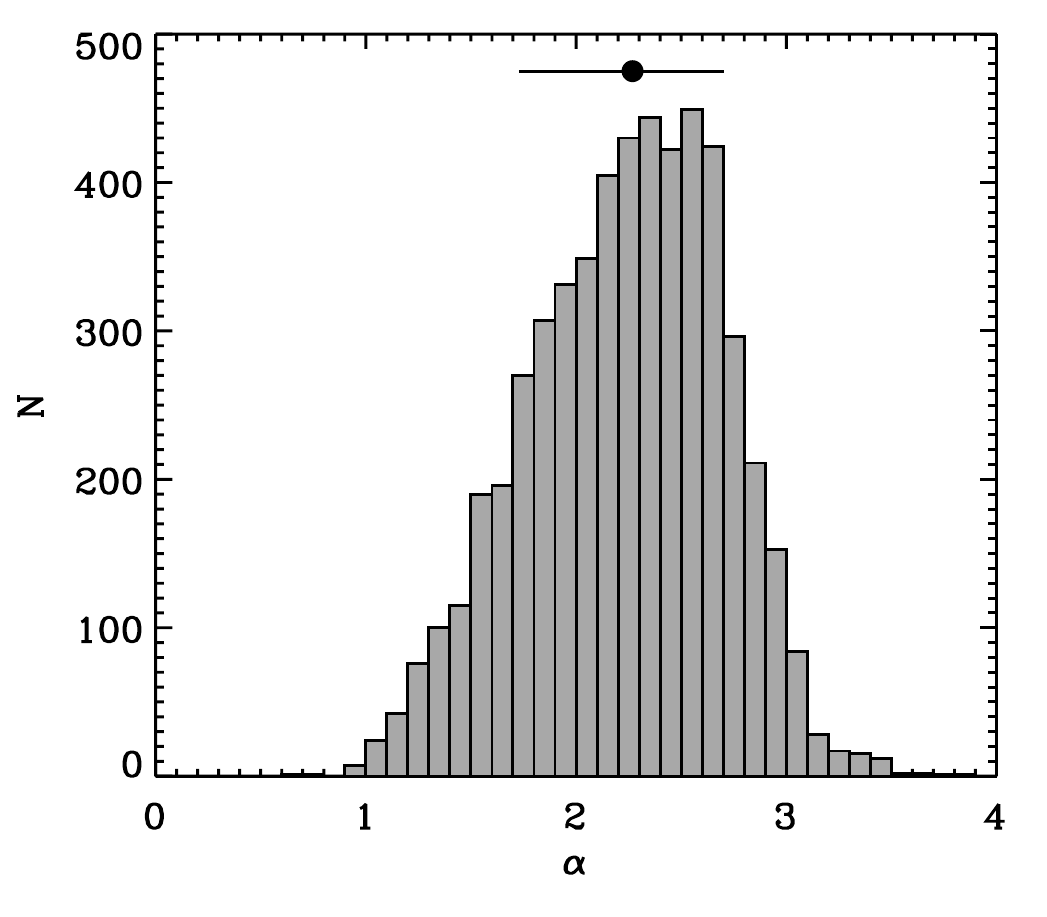}
\caption{\label{fg:alphadist} Distributions of the power-law index over the
  CGPS galactic plane data (i.e., excluding the high latitude extension).
  Note that the distribution of the index shows a significant skewness.   The
  bar above the distribution shows the median value and a line ranging from
  the 15th to 85th percentile of the distribution. }
\end{figure}

Figure \ref{fg:alphadist} shows the distribution of the spectral index over
the CGPS Galactic plane data.   The distribution appears non-Gaussian showing
significant negative skew.  We find that the median value of $\alpha$ is $2.26^{+0.44}_{-0.54}$ where the uncertainties are the difference between the median and the 15th and 85th percentiles.  If the distribution were normal, these would be the $\pm 1 \sigma$ points. Given the uncertainties in the index are significantly smaller than this spread, the changes in the shape of the power spectrum are significant.  The variations are also non-random, and appear correlated with the structure of the ISM in total intensity at 1420 MHz as well as with structure in other wavebands.  These correlations provide some insight as to the nature of the changing power spectrum.   The significant range in indices is inconsistent with assuming density fluctuations driven by a single Kolmogorov turbulent power spectrum for lines of sight in the Galactic plane \citep[e.g.,][]{minter96,have08}.  At high latitudes (Section \ref{sn:highlat}), a single value of $\alpha \sim 3$ may be an appropriate description of the emission structure.  

\subsection{Dependence on Stokes Parameters}
\label{sn:qucomp}

\begin{figure}
\plotone{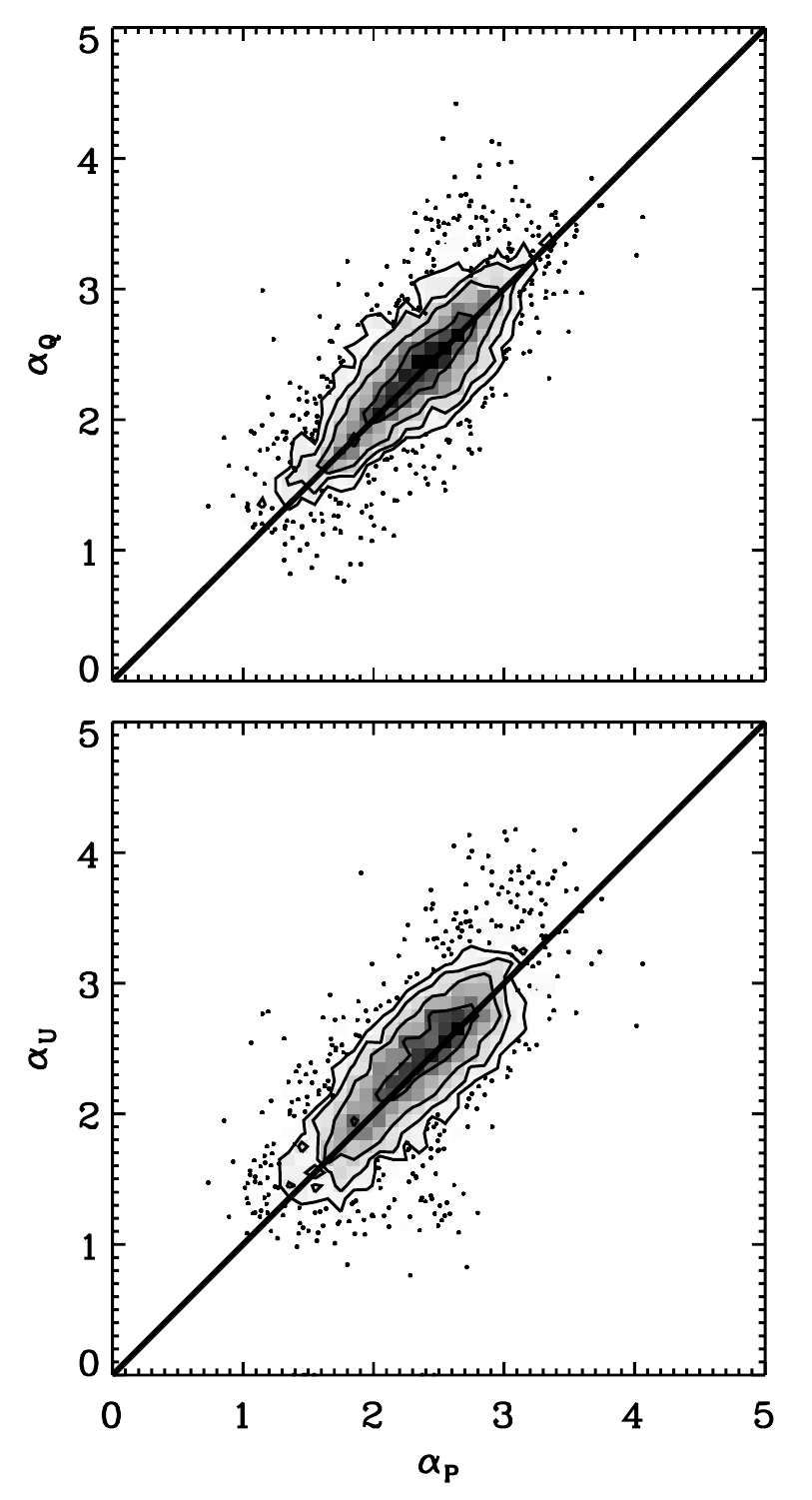}
\caption{\label{fg:qucomp} Comparison of the spectral index derived from analysis of the polarized intensity $\alpha_P$ to that derived from a parallel analysis of the Stokes $Q$ and $U$ parameters ($\alpha_Q$ and $\alpha_U$ respectively).  The different points represent spectral indices from the 5440 different submaps of the Galactic plane.  The grayscale and contours indicate the density of points where the plotting symbols overlap.  The black line indicates the locus of equality.  The  indices show good agreement among all three representations of the polarized emission. }
\end{figure}

We have presented power spectra for the polarized intensity $P=\sqrt{Q^2+U^2}$ but we have carried out a parallel analysis for Stokes $Q$ and $U$ data. In Figure \ref{fg:qucomp} we compare the power-law indices derived for $Q$ and $U$ with those derived for $P$.  There are 5440 data points but only $\sim 1/64$ of these are independent.  Overall, there is good agreement between the spectral index for polarized intensity $\alpha_P$ and that derived for the Stokes $Q$ and $U$ images: $\alpha_Q$ and $\alpha_U$.  The scatter is consistent with the uncertainties in the parameter estimates ($\sim 0.15$).  There is a small, marginally significant systematic offset between the spectral indices for the Stokes parameters and that for the polarized intensity: $\langle \alpha_P -\alpha_Q \rangle=-0.04\pm 0.03$ and $\langle \alpha_P-\alpha_U\rangle = -0.06\pm 0.03$.  The other parameters of the fit are also consistent between the three different images representing the polarized emission.  

We note that the structure in the Stokes $I$ maps is smooth compared to $P$, $Q$ and $U$ images, which implies that the amplitude fluctuations traced in the power spectrum arise from Faraday depolarization during propagation.  Both \citet{tucci02} and \citet{bruscoli02} find differences between the power spectra for individual components of the polarization (that work uses the equivalent description of $E$- and $B$-modes ).  In particular, they argue that the Stokes $Q$ and $U$ data would be sensitive to both changes in polarization angle and polarized intensity, whereas the $P$ data will only be sensitive to the polarized intensity.  \citet{tucci02} find that the indices for the $E$- and $B$-modes are larger by about 1.0 compared to the polarized intensity values in six separate regions at 1.4 GHz.  The primary difference bewteen their data and the CGPS data is the inclusion of large scale data in this work \citep[$\ell>600$ for ][]{tucci02}, and this difference may give rise to the apparent discrepancy in results.  Some regions in our analysis do show what appear to be significant differences between $\alpha_P,\alpha_U$ and $\alpha_Q$ and the wider area under consideration reduces the impact of effects from indivdiual regions.

The consistency of the $Q$, $U$ and $P$ power spectra give insight into the origin of structure in the polarized emission.  The $P$ power spectrum should be sensitive to amplitude fluctuations alone whereas $Q$ and $U$ should trace both amplitude and angle fluctuations \citep{tucci02}.   If polarization angle changes on scales larger than the beam, we should see significant differences between indices for the Stokes parameter maps and the polarized intensity maps.  Such changes can arise from a smooth Faraday screen and would affect $Q$ and $U$ but not $P$.  Such screens are seen in 5 GHz studies \citep{sun07,gao10} of the polarized continuum and may be affecting the 1.4 GHz emission along certain lines of sight.  These effects are likely small at 1.4 GHz since Faraday effects grow quadratically with increasing wavelength.  We find that the power spectra are overall quite similar for the $P$, $Q$ and $U$ images, which implies the structural changes are dominated by amplitude fluctuations instead of angle fluctuations.  From these images alone, we cannot distinguish between fluctuations in the synchrotron emission regions themselves (either in electron density or magnetic field) and Faraday depolarization effects coming from rotation measure fluctuations on scales smaller than a beam.  

We note that there is a small, systematic increase in $\alpha_{Q}$ and $\alpha_{U}$ with respect to $\alpha_{P}$ at $\alpha_P\sim 3$.  These data are mostly associated with a single region at $(75^{\circ},2.5^{\circ})$, near but not directly toward the Cygnus X region.  There are no obvious features in the H$\alpha$ maps or the Stokes $I$ image, suggesting this is a polarized intensity feature with small Faraday rotation.  A similar feature being present in the \cite{tucci02} data may explain the significant differences between indices derived for $\alpha_{Q},\alpha_{U}$ and $\alpha_P$ seen in that work.

\subsection{Changing Spectral Index with Latitude}
\label{sn:highlat}

\begin{figure}
\plotone{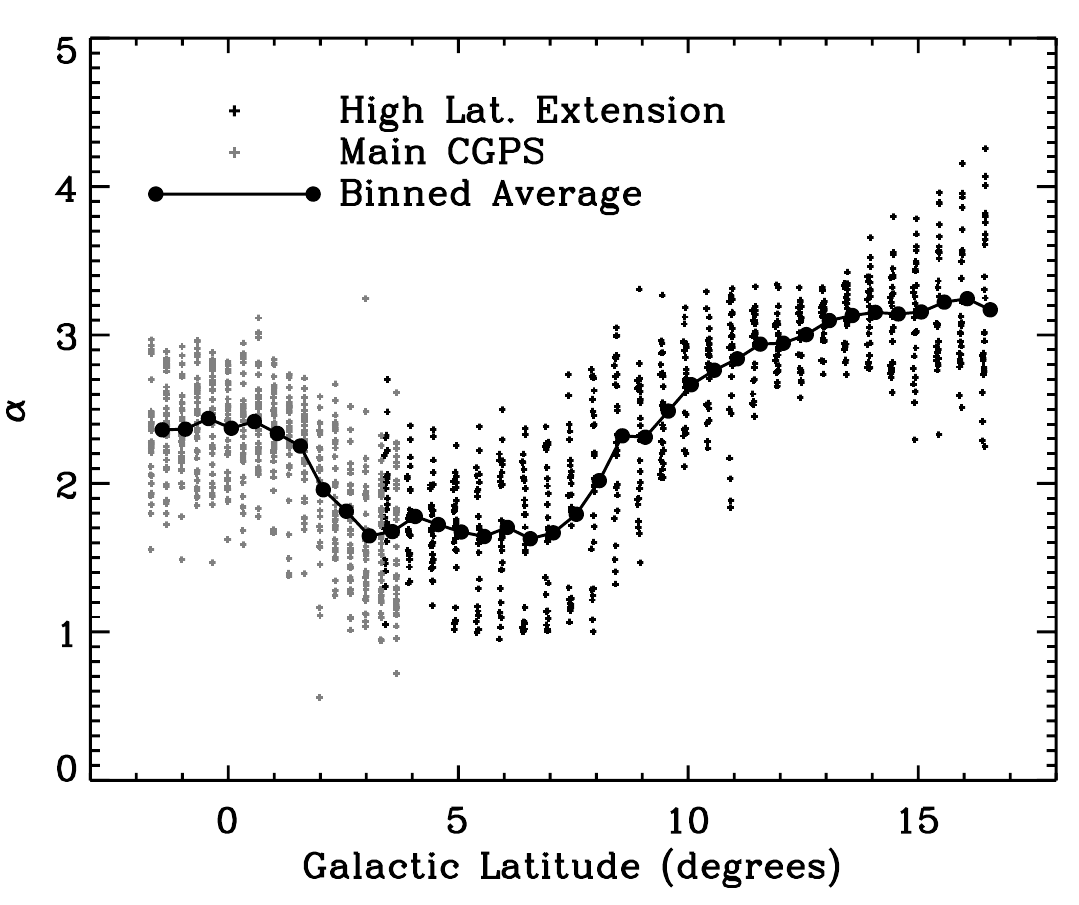}
\caption{\label{fg:alphab} Power-law index of the angular power
  spectrum as a function of Galactic latitude.  The main CGPS data span the entire survey range ($65^{\circ}\leq l\leq 175^{\circ}$) while the high latitude data span $101\degr \leq l \leq 116\degr$. }
\end{figure}

In Figure \ref{fg:alphab} we plot the spectral index vs. Galactic latitude; there is a significant variation with latitude. The most prominent variation is a gradual increase of spectral index across the high-latitude extension of the CGPS (in the range $101^{\circ} \leq  l \leq 116^{\circ}$).  The decrease in spectral index at ${b}\sim{2^{\circ}}$ is remarkably sharp, apparently unresolved by our $2.67^{\circ}$ window, which is surprising considering that the data are drawn from over $100^{\circ}$ of longitude.  The change at ${b}\sim{9^{\circ}}$ appears to be marginally resolved.

Compared to the Galactic plane distribution seen in Figure \ref{fg:alphadist}, the high latitude power spectrum is quite steep with $\langle \alpha\rangle > 3$.  This transition at $b\sim 9^{\circ}$ is likely associated with the change of the disk structure to that of the halo.  However, without knowing a characteristic depth into the Galaxy for the formation of polarized structure, this scale height cannot be determined unambiguously.   In the $15\degr$ range of longitudes where we have the high-latitude data, the transition appears at a roughly constant latitude; but this trend may vary significantly with longitude due to local features.

The sharp transition at $b\sim 2^{\circ}$ does not have a clear origin, but appears to hold on average across the entire CGPS survey range.  We conjecture that the steep power spectra arise toward the midplane because of the high number of ionized regions along those lines of sight.  These regions depolarize emission from more distant emitters, which would have smaller characteristic angular scales.  This depolarization would lead to less small scale structure and a steeper power spectrum.  Finding a similar change toward shallower indices at $b<-2^{\circ}$ would support this idea.

\citet{carr10} and \citet{have03} describe observations with a sufficiently broad range of latitudes to establish a trend. \citet{carr10} find a general increase of the spectral index $\alpha$ from $b = 0^{\circ}$ to $b \sim -60^{\circ}$, and a decrease of the spectral index from $b \sim -60^{\circ}$ to $b \sim -90^{\circ}$.  They also argue for a disk-halo transition at different latitudes ($|b|\sim 20^{\circ}$) than are observed in the CGPS data though their data differ in both frequency (2.3 GHz) and longitude ($\ell\sim 254^{\circ}$).  \citet{have03} finds a decrease in the spectral index with increasing Galactic latitude up to $b = 25^{\circ}$, however the latitudes below $b = 10^{\circ}$ are not well sampled, and there is no change from $b = 5^{\circ}$ to $10^{\circ}$. The results of the present investigation do not disagree with either of these previous studies, and the apparent discrepancy between $b = 10^{\circ}$ to $25^{\circ}$ in the other two studies may be attributable the different frequencies observed in those studies (350 MHz for Haverkorn et al., and 2.3 GHz for Carretti et al.). As the spectral indices seem to respond strongly to individual features that may be physically close, trends with Galactic latitude may be overwhelmed easily by local features.

\subsection{Large and small-scale features in the Galactic plane}
\label{sn:features}
Here, we briefly discuss characteristics of the spectral index maps and a comparison with data from the Canadian Galactic Plane Survey \citep{taylor03} and low resolution H$\alpha$ maps \citep{fink03}. The images are displayed in Figures~\ref{fg:maps} to \ref{fg:mapsend}. Over the entire area of the CGPS the polarized sky bears very little resemblance to the total-power sky \citep{landecker10}, a remark that can be found in every publication about large-scale polarization surveys at L-band and lower frequencies. The reason for this disagreement is that structure in the linear polarization images at low radio frequencies is largely produced by Faraday rotation rather than variation in the intrinsic synchrotron emission. Since we analyze power spectra of the linearly polarized emission, we expect similar behavior for the spectral index maps.

In general we should only receive linearly polarized emission which is generated between us and an imaginary border, which is referred to as the ``polarization horizon'' \citep{uyaniker03}. This is defined as the maximum distance from which we can still detect linearly polarized emission at radio frequencies. More distant emission is completely depolarized. This polarization horizon depends strongly on the resolution and observing frequency. \citet{kothes04} located the polarization horizon in a preliminary study of supernova remnants (SNRs) and \ion{H}{2} regions in the CGPS at 21~cm wavelength. They found a distance of about 2.5 kpc for longitudes below 140$^{\circ}$, while towards the anti-center it is likely much more distant.

Here, we try to relate features found in our spectral index maps to known nearby objects in the Galaxy like \ion{H}{2} regions or supernova remnants, which are located between us and the polarization horizon. Since our spectral index maps have a low resolution ($\sim 2.5\degr$), only very large objects will have a notable impact. We will also attempt to analyse spectral index features that do not have a counterpart in the CGPS 21-cm radio continuum images.  

\subsubsection{Supernova Remnants}

In the area of the CGPS discussed here, there are only 4 SNRs large enough to potentially have a noticeable impact on the polarization power spectra \citep{kothes06}.

\noindent{\bf CTB\,80 (G69.0$+$2.7)}: SNR CTB80 has a diameter of about $1.5\degr$ and is located at a distance of about 2.0 kpc \citep[][; in Figure~\ref{fg:mapsend}]{strom00}. The average spectral index in the region is $\alpha = 2.37 \pm 0.04$, slightly steeper than its surroundings.   Since the polarized emission in this region of the data is very low, the spectral index of the surrounding emission is dominated by noise. This may also affect the SNR's spectral index since its size is smaller than the resolution of our spectral index maps.

\noindent{\bf HB\,21 (G89.0$+$4.7)}: SNR HB~21 is located at the upper edge of the CGPS and is not fully covered in this analysis.  SNR HB~21 has a diameter of about $2\degr$ and is located at a distance of 800~pc \citep{tatematsu90}. We find a spectral index of $\alpha = 2.09 \pm 0.24$ (see Figure~\ref{fg:mapse}). 
Since little of HB\,21 is seen in the index maps and we find both steeper and flatter spectra in the surrounding area, we cannot rigorously compare the SNR to the surrounding region.

\noindent{\bf CTB\,104a (G93.7$-$0.2)}: The local SNR CTB104a has a diameter of about $1.5\degr$ and is located at a distance of 1.5~kpc \citep{uyaniker02}. We find a spectral index of $\alpha = 1.74 \pm 0.15$ (see Figure~\ref{fg:mapse}). The SNR dominates the polarized emission in this part of the sky and we consider this value to be a reliable description of the SNR emission.

\noindent{\bf HB\,9 (G160.9$+$2.6)}:  With a diameter of about $2.5\degr$, HB\,9 is the largest SNR covered by the area of the CGPS discussed here. \citet{lozinskaya81} found a radial velocity of about $-18$~km\,s$^{-1}$ and a distance of 2~kpc, which would make this SNR a resident of the Perseus spiral arm.  HB~9 is the largest of the 4 SNRs and shows an unmistakable signal (see Figure~\ref{fg:maps}) with spectral index $\alpha = 1.86 \pm 0.17$, very similar to CTB\,104a.

The major emission mechanism for radio emission in supernova remnants is synchrotron emission and therefore these objects are emitters of linearly polarized emission. This polarized emission should be intrinsically smooth with the same power spectrum as the total power emission. Small-scale fluctuations in the appearance of the polarized emission are introduced by Faraday rotation, which arises either internally from magnetic fields and thermal electrons in the SNR itself or in the foreground.  A comparison of the SNR's power spectrum with that of its environment should indicate whether internal or foreground Faraday rotation is the dominant factor.

Our results for the SNRs CTB\,80 and HB\,21 are inconclusive: CTB\,80 is likely too small and has a noisy measurement of the surrounding environment, and HB\,21 is not fully covered by our observations. The area around HB\,9 displays significantly steeper power spectra than the SNR. Therefore the Faraday rotation producing the flat power spectrum is most likely internal to HB\,9 and not in the foreground, because otherwise the surroundings of HB\,9, which show very bright polarized emission, should have a flat power spectrum as well. On the other hand the spectral index we find for CTB\,104a is very similar to its surroundings.  For this SNR, this points to the Faraday rotation being predominantly in the foreground.

We have succeded in detecting the effects of two SNRs in our data.  While we have suggested a framework for interpretation of such data, power spectra from observations of higher angular resolution are needed to make a more thorough study of this topic.

\subsubsection{\ion{H}{2} regions}
\label{sn:hii-regions}
There are a number of nearby \ion{H}{2} regions and \ion{H}{2} region complexes large enough for us to investigate.

\noindent{\bf The Cygnus X region} is a complex region in the Galactic plane where the line of sight runs tangent to the local spiral arm, the Orion Spur ($76.5^{\circ}<l<83^{\circ}, -1^{\circ}<b<4^{\circ}$). Cygnus X consists of many layers of objects at various distances, the closest of which is the dark cloud and dust complex called the Great Cygnus Rift.  The Rift  should not completely depolarize its background but may cause some Faraday rotation. There are star forming and \ion{H}{2} region complexes related to W75N and DR21 \citep{gottschalk12}, however, which are located at 1.3 and 1.5 kpc, respectively \citep{rygl12}. The ionized nebulae related to W75N and DR21 should depolarize any linearly polarized emission from behind them, which gives us the opportunity to study synchrotron emission that is generated and Faraday rotated in the intervening 1.3 kpc. The Cygnus X region has a relatively flat spectral index compared to its surroundings, which show steep and very noisy spectra (see Figure~4f).  There is strong depolarization of background emission by the thermal gas in Cygnus X, so that the detected structure arises from foreground emission and Faraday rotation.  The spectral index twards Cygnus X is $\alpha = 2.25 \pm 0.15$ with only small variation over the entire complex

\noindent {\bf The W~80 \ion{H}{2} region complex} (G$85.0-0.8$) is found adjacent to Cygnus~X. It contains two ionized nebulae: the Pelican Nebula and the North America Nebula. This ionized gas is at a distance of about 800~pc \citep{blitz82}. W~80 seems to display a somewhat steeper power spectrum than Cygnus~X. There is suggestion of a gradient across W~80 from steeper spectra at lower longitudes to shallower at higher longitudes. However, this may be the result of the ambient power spectra that become more dominant at the side of the region. In the center of W~80, the spectral index is $\alpha = 2.65 \pm 0.12$.

\noindent {\bf The \ion{H}{2} region Sh 2-124} (G$94.6-1.5$), adjacent to the SNR CTB\,104a, has a diameter of about $1\degr$ with a compact core.  There is no obvious corresponding signature in the spectral index map, which most likely indicates that this \ion{H}{2} region -- at a distance of 2.6~kpc \citep{blitz82} -- is located beyond the polarization horizon. This agrees with the results by \citet{kothes04}.

\noindent The nearby {\bf \ion{H}{2} region Sh~2-131} (G$99.3+3.5$) shows up as a prominent feature with a steep spectral index toward the lower part of the \ion{H}{2} region.  The polarized intensity map shows large-scale but faint structure across the region, consistent with the region depolarizing the background. There is significant power in the diffuse emission around the \ion{H}{2} region.  The signal from this \ion{H}{2} region is only clear where the $2.67^{\circ}$ window is aligned with the region.  The average spectral index is $\alpha = 2.85 \pm 0.28$ with a peak of $\alpha = 3.4$ near the centre. This \ion{H}{2} region has a diameter of $2.5\degr$, the largest of our sample at a distance of 900~pc \citep{humphreys78}.

\noindent {\bf W3/4/5} is a large \ion{H}{2} region complex in the Perseus arm ($132^{\circ} < l < 138.5^{\circ}, -0.5^{\circ}<b<2.0^{\circ}$) at a distance of 2.0~kpc \citep{xu06,hach06}. The polarized emission appears to be affected by the \ion{H}{2} region across a large area, even beyond the main thermal emission components.  However, the complex is not well distinguished from the background in the power-law maps.  We use only the eastern half to determine a mean spectral index of $\alpha = 1.76 \pm 0.17$, because the western half is strongly affected by artifacts.

\noindent In the region $172^{\circ}<l<175^{\circ}, -3^{\circ}<b<4^{\circ}$, we find a \ion{H}{2} region complex that contains the {\bf \ion{H}{2} regions Sh 2-229 to 2-237}. Some of these \ion{H}{2} regions are local and some are Perseus arm residents \citep{blitz82}. There is also diffuse thermal radio emission in between. The spectral index in this area is rapidly fluctuating, probably because of the variation in distance. We find two areas, at the top and bottom, which have steep power spectra, divided by a flat spectrum region in the middle. Since most of the \ion{H}{2} regions are concentrated at the top and bottom of the complex, we can assume that those are completely depolarizing their background, producing steep spectra, and the center might be partly transparent to background polarization.

\ion{H}{2} regions in the Canadian Galactic Plane Survey give us the opportunity to study synchrotron emission and Faraday rotation which are produced nearby, since they depolarize all linearly polarized emission coming from behind them, because of high electron densities and small-scale turbulence \citep{landecker10}. Steeper power spectra should indicate a closer distance, assuming similar cell-sizes for turbulent structures. The relatively constant polarization horizon over a large portion of the CGPS indicates that this assumption is valid.

We have obtained reliable spectral indices for four \ion{H}{2} regions or
complexes of \ion{H}{2} regions.  There is a tendency for closer objects to
have steeper spectra. The W3/4/5 \ion{H}{2} region complex is at the largest distance and shows the flattest power spectrum and W80 and Sh 2-131 are closest and have the steepest spectra with Cygnus X in between. However, for a more thorough analysis we would need to analyze data for more \ion{H}{2} regions, which requires power spectra based on observations with higher angular resolution.

\subsubsection{Other Spectral Index Features}

Synchrotron emission generated in our Galaxy from cosmic-ray electrons
interacting with the large-scale magnetic field should be intrinsically
very smooth. We find typically a steep spectral index $\alpha$ for
presumably ``untouched'' synchrotron emission features. This can be found in many areas of the CGPS data set, in particular towards the so-called Fan-region,
a highly linearly polarized feature that extends above the mid-plane from a 
Galactic Longitude of $120^\circ$ to just beyond the 
anti-center \citep[see Figure 11 in][]{wolleben06}. Areas of this kind
can be seen in Figs.~\ref{fg:maps} to \ref{fg:mapsc}. 

There is a molecular cloud complex next to HB~21 at higher longitudes
with which the SNR is probably interacting \citep[G$92.0+3.5$;][]{tatematsu90}. This molecular cloud complex coincides spatially with an area of very flat
power spectra. The resulting spectral index is $\alpha = 1.44 \pm 0.21$,
which is the lowest value for $\alpha$ we found for a distinct object.
It is not unusual that a molecular cloud complex imposes Faraday rotation on its
polarized background. A very thorough study of Faraday rotation related to
the Taurus molecular cloud complex was presented by \citet{wolleben04},
who found highly structured polarized emission.  Their simulations
suggest the structure arises from Faraday rotation features, caused mainly by very strong magnetic fields related to those molecular clouds.

There are many spectral index features in Figures~\ref{fg:maps} to
\ref{fg:mapsend} that we cannot relate to any known object visible
in radio continuum or H$\alpha$. These are probably areas where the
electron density and/or magnetic field are so subtly enhanced that we cannot detect them as emission features.  Between Galactic longitude of about $94\degr$ and $99\degr$ and Galactic latitude of about $0\degr$ and the edge of the CGPS, there is a feature of a distinct power spectrum -- steeper than its surroundings -- that can be identified in the polarization images as an area of significant structure in polarized intensity. This area seems to coincide with a gap in H$\alpha$ emission. On the other hand, there are areas of enhanced H$\alpha$ emission between Galactic longitude of $148\degr$ and $153\degr$ that coincide with spectral index features as well. In this case, however, the power spectra are flatter than their environment.

\subsection{The Effect of Thermal Emission}

The turbulent structure of the magnetoionic medium leads to the observed structure in the polarized emission.  In parameterizing this structure with a power law, we are describing whether the structure is found on small or large scales.  However, the power-law component as a whole provides a good measure of the emission from this phase of the medium.  By separating the structure in Fourier space from other effects, we can examine how this component correlates with other phases of the ISM. 

We establish the power in the power-law component of the fit through the fit parameters, integrating 
\begin{equation}
W(B,\alpha) = \int_{\ell=30}^{\ell=10^4} B\ell^{-\alpha}\, 2\pi \ell\, d\ell.
\end{equation}
Here the bounds of the multipoles are selected to cover the
range well sampled by the CGPS data.  This integrated power is robust against
the effects of the window function, the covariance between the fit parameters, and the presence of point sources in the data.  Thus, we prefer this as a representation of the power-law component of the images to using the polarized intensity directly.  

\begin{figure}
\plotone{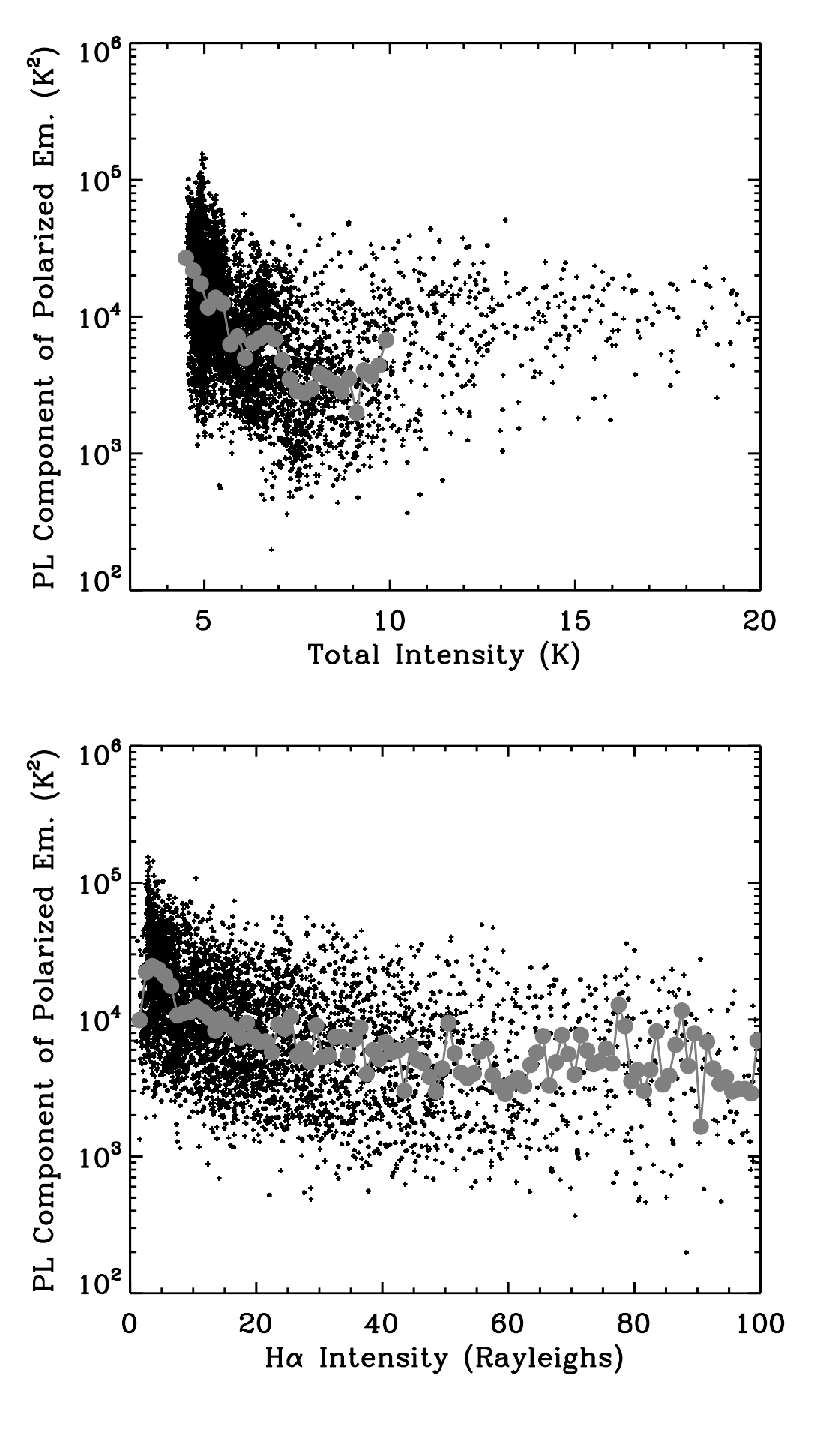}
\caption{\label{fg:anticorr} Integrated power in the power-law component of
  the power spectrum as a function of total 1420 MHz intensity (Stokes $I$, top) and H$\alpha$ emission (bottom).  With the exception of the brightest total intensity data attributable to point sources, the power-law component of the polarized intensity declines with total intensity surface brightness.  A similar trend is seen in the H$\alpha$ emission.  Gray points indicate binned averages to represent the mean trend of the data.}
\end{figure}

In addition to polarized intensity information, the CGPS also maps thermal emission from the warm ionized medium.  We compare our results to the CGPS $I$ images, which contain a significant fraction of thermal emission, and to the mosaic of H$\alpha$ emission from \citep{fink03}.  We identify an anticorrelation of the polarized intensity with both the Stokes $I$ and H$\alpha$ data as shown in Figure \ref{fg:anticorr}.   For the H$\alpha$ data, the anticorrelation should represent a wide-area probe of the Faraday depolarization by the warm ionized medium. The high electron densities implied by bright thermal emission eliminate structure behind the emission, reducing the resultant power and steepening the spectrum (Section \ref{sn:hii-regions}).  At lower brightness, the thermal emission may indicate regions that will not completely depolarize the emission behind them, and merely alter the shape of the spectrum.  It is possible that the depolarization of background emission would enhance the amount of image power present in the structured (power-law) component. For the Stokes $I$ data, the interpretation of the anticorrelation is less clear since the waveband includes both thermal and nonthermal (intrinsically polarized) emission.  Along low-latitude lines of sight where the Stokes $I$ emission is brightest, the nonthermal emission may also be depolarized by the superposition of many polarized signals (depth depolarization). 

\section{Conclusion}
We present a spatial power spectrum analysis of the polarized emission from the Galactic plane as mapped in the CGPS by \citet{landecker10}.  The CGPS is unrivalled in its combination of Galactic longitude coverage, angular resolution, and full inclusion of a wide range of angular scales.  This power spectrum analysis is the largest such analysis to date, and we are able to resolve significant spatial variations in the index of the power spectrum at a resolution of 2.67$^{\circ}$.  We have examined instrumental effects and spurious contributing signals and developed a model which accounts for these contaminants.  

We find significant, large-scale variations in the slope of the power spectrum associated with different features of the interstellar medium.  There are pronounced shifts toward large-scale structure in the direction of \ion{H}{2} regions, which are likely the result of the Faraday rotation eliminating polarized signal from behind the ionized region.  We find no difference between the spatial power spectrum index calculated for the Stokes $Q$ and $U$ data and that calculated for the polarized intensity.  We interpret this similarity as evidence that the structure in the 1.4 GHz polarized emission arises mostly from Faraday depolarization effects.   Lines of sight with large Faraday depths have contributions from many different regions and the emission is Faraday rotated in transmission, leading to a change in the structure.  Additionally, there are sharp changes in the slope of the power spectrum at high Galactic latitudes.  These changes can represent both a change in the structure of the emitting medium and short lines of sight through the disk arising above $b>10^{\circ}$ in the 15$^{\circ}$ longitude range centered at $l\sim 108^{\circ}$.  Finally, we note an anticorrelation between the brightness of the diffuse polarized emission and the Stokes $I$ 21-cm continuum emission as well as the H$\alpha$ emission.

The polarized sky still holds much to be discovered, and a full classification of the features in the polarized sky may soon be possible as large surveys such as the CGPS reveal polarization structure on all scales. Decoupling small-scale magnetic field variation and electron densities from the regular component of the magnetic field remains difficult at 1.4 GHz.  Recent progress in parameterizing the resulting emission properties from generalized magnetohydrodynamic turbulence theory \citep[e.g.,][]{lp12} offers some possibility for interpreting such data.  However, the theoretical framework currently assumes constant electron density and will need generalization to include fluctuations in electron density to capture Faraday rotation effects and thus be directly applicable to these data.  The results of our angular power spectrum analysis find real features captured in the structure of the polarized emission, and we expect that these results can soon be translated into a characterization of the turbulent magneto-ionic medium.

\acknowledgements
The research presented in this paper has used data from the Canadian Galactic Plane Survey, a Canadian project with international partners, supported by the Natural Sciences and Engineering Research Council (NSERC) Canada.  The authors of the paper are all supported by Discovery Grants from NSERC.  This research made use of NASA's Astrophysics Data System.  We thank Marijke Haverkorn for useful discussions that improved the quality of this paper.  The manuscript benefited from the constructive comments of the referee Ettore Carretti (CSIRO), in particular regarding connections to previous work and the utility of the various parameterizations of the polarized emission in a power spectrum analysis.

{\it Facilities:} \facility{DRAO: Synthesis (1.4 GHz)}
\bibliographystyle{apj}
\bibliography{ms}

\end{document}